\documentclass[pra,showpacs]{revtex4}
\newcommand{\dket}[1]{| \, #1 \rangle\!\rangle}
\newcommand{\dbra}[1]{\langle\!\langle #1 \, |}
\newcommand{\N}[1]{|\!| #1 |\!|}
\def\kk{\rangle\!\rangle}
\def\bb{\langle\!\langle}
\def\leq{\leqslant}
\def\>{\rangle}
\def\<{\langle}
\def\d{\mathrm d}
\def\sH{{\cal H}}
\def\sK{{\cal K}}
\def\Tr{\hbox{Tr}}
\newcommand{\SU}[1]{\mathbb{SU}(#1)}
\usepackage{amsmath,amssymb}
\begin{document}

\title{Informationally complete measurements on bipartite quantum
  systems: comparing local with global measurements} \author{G. M.
  D'Ariano, P. Perinotti, and M. F. Sacchi} \affiliation{QUIT, Unit\`a
  INFM and Dipartimento di Fisica ``A. Volta'', Universit\`a di Pavia,
  I-27100 Pavia, Italy.} 
\date{\today}

\begin{abstract}
  Informationally complete measurements allow the estimation of
  expectation values of any operator on a quantum system, by changing
  only the data-processing of the measurement outcomes. In particular,
  an informationally complete measurement can be used to perform
  quantum tomography, namely to estimate the density matrix of the
  quantum state. The data-processing is generally nonunique, and can
  be optimized according to a given criterion. In this paper we
  provide the solution of the optimization problem which minimizes the
  variance in the estimation. We then consider informationally
  complete measurements performed over bipartite quantum systems
  focusing attention on universally covariant measurements, and
  compare their statistical efficiency when performed either locally
  or globally on the two systems.  Among global measurements we
  consider the special case of Bell measurements, which allow to
  estimate the expectation of a restricted class of operators. We
  compare the variance in the three cases: local, Bell, and
  unrestricted global---and derive conditions for the operators to be
  estimated such that one type of measurement is more efficient than
  the other. In particular, we find that for factorized operators and
  Bell projectors the Bell measurement always performs better than the
  unrestricted global measurement, which in turn outperforms the local
  one. For estimation of the matrix elements of the density operator,
  the relative performances depend on the basis on which the state is
  represented, and on the matrix element being diagonal or
  off-diagonal, however, with the global unrestricted measurement
  generally performing better than the local one.
\end{abstract}

\maketitle

\section{Introduction}
One of the most distinctive features of quantum mechanics is that the
information on the state of the quantum system cannot be retrieved in
a single measurement \cite{impo}. Among the different methods for
retrieving information on the state---including quantum tomography,
state discrimination, and quantum state estimation \cite{sv}---the
informationally complete measurement \cite{prug,bus,univest} (shortly
named {\em infocomplete} in the following) is the most versatile,
allowing one to obtain the expectation $\<O\>$ of any operator $O$ of
the system, and with the additional possibility of adapting the
processing of outcomes, depending on the quantity to be estimated and
on the set of states on which the measurement is performed. A
particularly interesting case is when the measurement is {\em
covariant} \cite{Holevo,ourcov} with respect to a group of physical
transformations. This means that there is an action of a group on the
probability space which maps events into events in a covariant
fashion, namely that when the quantum system is group-transformed, the
probability of the given event becomes the probability of the
transformed event. This situation is very natural, and occurs in many
applications (e~.g. heterodyne detection \cite{Bilk-poms}, measurement
of the spin-direction \cite{spin}, transmission of reference frames
\cite{refframe}, estimation of unitary transformations
\cite{EntEstimation}, etc.)

In this paper we study covariant infocomplete measurements for
bipartite quantum systems. The problem of optimization of
data-processing will be solved using the method of {\em frame theory}
\cite{ds,czz}, which recently has been shown \cite{ic1,ic2} to be
perfectly suited to optimize covariant infocomplete measurements.
Dealing with bipartite (or multipartite) quantum system, the study of
the possibility to perform information-theoretic tasks locally on
non-local states has become a new paradigm in the field of quantum
information. In this respect consider for example the study of local
implementation of non-local quantum gates \cite{gates}, telecloning
\cite{Murao 99}, remote manipulation and preparation of quantum states
\cite{remote}, and local discrimination of non-local states
\cite{locstat}. The problem of comparing the performances of local and
entangled estimation schemes has not been addressed yet, and this is
precisely the purpose of the present paper, where we compare the
statistical efficiency of local infocomplete measurements with that of
nonlocal ones. In particular, the possibility of having a maximally
entangled infocomplete measurement---or {\em Bell quorum}---has never
been considered. Here we will show that, although an infocomplete Bell
measurement strictly does not exist, however, Bell measurements can be
used to estimate the expectation of any operator $O$ having both
partial traces $\Tr_1[O]$ and $\Tr_2[O]$ proportional to the identity,
e.~g.  maximally entangled operators and traceless factorized
observables.  This possibility is interesting also in view of the
relevant problem of the classification of measurements according to
separability criteria, which still remains unsolved.

The statistical efficiency of an infocomplete measurement is
quantified by the noise in the estimation of the operator expectation
$\<O\>$.  This generally depends both on the state and on the operator
$O$ to be estimated. However, for unbiased estimations the squared
expectation is independent of the measurement, whereas if the
measurement is also fully covariant (i.~e.  covariant under the full
group of unitary transformations) the square of the estimation
averaged over the ensemble of possible input states depends only on
the measurement: this greatly simplifies the evaluation of the
variances, and allows a simple comparison of statistical efficiencies
of different infocomplete measurements.

\par The paper is organized as follows. After summarizing some useful
notation and elements of the {\em Theory of Frames} in Section
\ref{s:frame}, in Sect. \ref{s:processing} we give a general method
for optimizing the data-processing. In Sect. \ref{s:cov} we consider
covariant infocomplete measurements for bipartite quantum systems, in
which case the optimal data-processing is provided by the so-called
{\em canonical dual frame}.  In Sections \ref{s:factorized},
\ref{s:global}, and \ref{s:Bell} we derive the minimal noise (averaged
over an ensembles of covariant states) for factorized, global, and
Bell infocomplete covariant measurements, and finally we conclude the
paper in Section \ref{s:comparison} by comparing the resulting
statistical efficiencies of the different types of measurements.

\section{Notation and elements of the theory of frames}\label{s:frame}
An infocomplete measurement is characterized by a Positive
Operator-Valued Measurement (POVM) that spans the whole space of
linear operators. The {\em Theory of frames} \cite{ds,czz} provides
simple and powerful tools to establish whether a set of operators is
complete in the whole operator space, and in addition gives a simple
algebraic rule for constructing all possible expansion
coefficients---representing all possible data-processing of
measurement outcomes---in terms of the so-called {\em dual
frames}. Upon denoting with $\{P_i \}$ the elements of an infocomplete
POVM \cite{ic1}, dual frames correspond to sets of operators $\{D_i\}$
in terms of which we can write the operator expansion as follows
\begin{equation}
\sum_ {i=1}^N \Tr[D^\dag_i O]\, P_i=O\,,
\label{eq:reconst}
\end{equation}
where $N$ denotes the number of outcomes of the POVM (in the following
we will not specify it anymore in the sum limits).  When considering
expansions of operators over a Hilbert space $\cal H$ it is convenient
to exploit the natural isomorphism between operators $O$ on $\cal H$
and vectors $\dket{O}$ in ${\cal H}^{\otimes 2}$, defined through the
equation
\begin{equation}
\dket{O}\doteq\sum_{m,n}\<m|O|n\>|m\>|n\>\,.\label{tre}
\end{equation}
We will make repeated use of the following identities \cite{pla}
\begin{eqnarray}
&&A\otimes B\dket{C}=\dket{ACB^\tau}\,,\\
&& \Tr_1[\dket{A}\dbra{B}]=A^\tau B^*\,,\\
&& \Tr_2[\dket{A}\dbra{B}]=AB^\dag\,,
\end{eqnarray}
where $\tau $ and $*$ denote transposition and complex conjugation
with respect to the given fixed basis in Eq. (\ref{tre}). The main
results from frame theory are the following \cite{ic1}. First, a the
POVM $\{P_i \}$ is infocomplete if and only if the {\em frame
operator}
\begin{equation}
F=\sum _i \dket{P_i}\dbra{P_i}\,,
\end{equation}
is invertible. A dual frame $\{D_i \}$ is given by the following one (usually referred to as {\em
  canonical})
\begin{equation}
\dket{D_i}=F^{-1}\dket{P_i}\,.
\end{equation}
In general there exist infinitely many alternate duals $\{D'_i\}$ ,
which can be obtained from the canonical one as follows
\begin{equation}
D^\prime_i=D_i+Y_i-\sum _ j \Tr[P^\dag_j D_i]\,Y_j\,,
\end{equation}
for arbitrary set $\{Y_j \}$. Finally, whenever a POVM is not
complete, the span of its elements coincides with the {\em support}
(the orthogonal complement of the kernel) of the operator $F$. Then
the operators in such a subspace can be reconstructed through Eq.
(\ref{eq:reconst}), where the canonical dual is now defined through
the Moore-Penrose generalized inverse $F^\ddag$ of $F$, which
corresponds to invert $F$ on its support.

\section{Optimization of the data-processing}\label{s:processing}
In the following we will restrict our attention only to finite
dimension $d<\infty$ of the system Hilbert space.  Apart from the case
of POVM $\{ P_i\}$ with exactly $d^2$ linearly independent elements
$P_i$ (whence with discrete sample space), for infocomplete POVM's
there always exist infinitely many alternate duals.  This feature
provides a wide freedom in choosing the dual frame, which can be
exploited in order to optimize the data-processing, e.~g. to minimize
the variance in the estimation. Therefore, suppose that an
experimenter is collecting statistics using an infocomplete measuring
apparatus---e.~g. in a quantum tomographic setup---with the aim of
estimating the expectations of any desired operator. The noise in the
estimation of the operator $O$ can be evaluated by \cite{corr}
\begin{equation}
\delta O^2 (\rho ;P)
\doteq\sum _ i |\Tr[D^\dag_i O]|^2\Tr[P_i\rho]- |\Tr[O\rho]|^2\,,
\label{deltorho}
\end{equation}
which clearly depends on the state $\rho$. Notice that we consider
generally complex operators $O$, i.~e. we can include the case of
external products $O=|i\>\<j|$ of orthonormal basis, which are needed
to estimate the matrix elements $\rho_{ij}$ of the density operator
$\rho$. In a Bayesian scheme the noise (\ref{deltorho}) can be
minimized for a given prior probability distribution $ d\mu (\rho )$
of the input states $\rho$, corresponding to the prior state $\bar
\rho = \int d\mu (\rho )\rho$.  The dual frame $\{D_i\}$ can be chosen
in order to minimize such a noise, which in turn corresponds to
minimize the following norm
\begin{equation}
\frac1{d^2}\N{c}_\pi^2=\frac1{d^2}\sum_i |\Tr[D^\dag_i O]|^2\Tr[P_i
\rho ]\,,
\end{equation}
where $|c \rangle = \sum _{i=1} ^N \Tr[D^\dag_i O] |e_i \rangle $ is a
vector in ${\mathbb C}^N$ , and $\{e_i\}$ denotes the canonical basis
of ${\mathbb C}^N$. We will now state a useful condition for
optimality. Let us consider the linear mapping $\Lambda:{\mathbb
C}^N\to\sH^{\otimes2}$
\begin{equation}
\Lambda |c\>=\sum_ic_i |P_i\kk \,,
\end{equation}
where $c_i=\<e_i|c\>$. The matrix elements
representing $\Lambda$ in the bases $|e_i\>\in{\mathbb C}^N$ and
$|m\>|n\>\in\sH^{\otimes2}$ are given by
\begin{equation}\label{Lambda}
\Lambda_{mn,i}=(P_i)_{mn}\,.
\end{equation}
We can define $\Gamma$ to be a {\em generalized inverse} of $\Lambda$,
shortly g-inverse, if $\Lambda=\Lambda\Gamma\Lambda$ holds (in the
following, for all properties of the various kinds of g-inverse see
Ref. \cite{bapat}). We now show that any g-inverse $\Gamma$ must have
matrix elements of the form
\begin{equation}\label{Gamma}
\Gamma_{i,mn}=(D^*_i)_{mn}\,,
\end{equation}
where $\{D_i\}$ is a dual frame for $\{P_i\}$. In fact,
$(\Gamma\Lambda)_{ij}=\Tr[D^\dag_iP_j]$, and consequently the
condition for $\Gamma$ to be a g-inverse can be written as
$(\Lambda\Gamma\Lambda)_{mn,j}=\sum_i(P_i)_{mn}\Tr[D^\dag_iP_j]$.
Then, by informational completeness of the POVM $\{P_j\}$, the
reconstruction formula $\sum_i\Tr[D^\dag_i O]P_i=O$ must hold, and
necessarily $\{D_i\}$ is a dual. Moreover, since $\{D_i\}$ is a dual,
than the g-inverse is also {\em reflexive} , namely
$\Gamma\Lambda\Gamma=\Gamma$, as can be easily checked from the
expansion
$(\Gamma\Lambda\Gamma)_{i,mn}\equiv\sum_j\Tr[D_iP_j](D_j^\dag)_{nm}=
(D_i^*)_{mn}\equiv (\Gamma)_{i,mn}$. In addition, the g-inverse
$\Gamma$ is also a {\em least square} inverse, since $\N{\Lambda\Gamma
O-O}=0$, as a consequence of the expansion $(\Lambda\Gamma
O)_{mn}\equiv \sum_i\Tr[D_i^\dag O](P_i)_{mn}=(O)_{mn}$. Therefore,
summarizing, any dual frame $\{D_i\}$ corresponds to a reflexive and
least square g-inverse $\Gamma$ of $\Lambda$ through
Eqs. (\ref{Lambda}) and (\ref{Gamma}).  \par Now, the quantity we want
to minimize is $\N{\Gamma O}_\pi$, where the norm $\N\bullet_\pi$ is
defined through
\begin{equation}
\N c_\pi^2=\<c|\pi|c\>\,,
\end{equation}
$\pi$ being the diagonal matrix (in the canonical basis $e_i$) with
entries $\pi_{ii}=\Tr[P_i \rho ]$ (for finite dimension the POVM is
trace-class). Notice that when $\Tr[P_i \rho ]$ does not depend on $i$
(as in the case where the average over the prior distribution of
states gives $\bar \rho \propto I$ and the POVM is covariant so that
$\Tr[P_i \bar \rho ] \propto \Tr[U_i\nu U_i^\dag]=\Tr[\nu]$) then
$\pi\propto I$. The quantity to be minimized is then simply $\N
c^2=\<c|c\>$, and the solution to the minimum-norm problem is provided
by any matrix $\Gamma$ such that
$\Gamma\Lambda=\Lambda^\dag\Gamma^\dag=\Lambda^\dag\Gamma^\dag\Gamma\Lambda$.
Along with the reflexivity and least-square properties, the
minimum-norm condition uniquely determines the g-inverse in terms of
the Moore-Penrose pseudo inverse $\Gamma\equiv\Lambda^\ddag$. Since
$(\Gamma\Lambda)_{ij}=\Tr[D^\dag_i P_j]$, the previous identity is
equivalent to
\begin{equation}
\bb D_i|P_j\kk=\bb P_i|D_j\kk=\sum_k\bb P_i|D_k\kk\bb D_k|P_j\kk\,.
\end{equation}
Now it is easy to check that the canonical dual $D_i$, defined as
$|D_i\kk=F^{-1}|P_i\kk$, satisfies the previous condition, as
$\sum_k|D_k\kk\bb D_k|=F^{-1}$, and
\begin{equation}
\bb D_i|P_j\kk=\bb P_i|D_j\kk=\sum_k\bb P_i
|D_k\kk\bb D_k|P_j\kk=\bb P_i|F^{-1}|P_j\kk\,.
\end{equation}
In the general case in which $\pi\not\propto I$, analogous proof as in
the previous case \cite{bapat} leads to the following condition for
minimization of $\N{\Gamma O}_\pi$
\begin{equation}
\pi\Gamma\Lambda=
\Lambda^\dag\Gamma^\dag\pi=\Lambda^\dag\Gamma^\dag\pi\Gamma\Lambda\,.
\end{equation}
The results of the present and the previous sections also apply in the
case of continuous POVM by suitably replacing the discrete index $i$
with a continuous one, and sums with integrals with some care about
summability of integrals in the case of noncompact groups.

\section{Infocomplete covariant measurements for bipartite systems}
\label{s:cov}
In the following we will consider quantum measurements that are
covariant under the action of a group ${\bf G}$ of unitary operators
$U_g$, $g\in{\bf G}$, on the Hilbert space ${\cal H}^{\otimes2}$ of a
bipartite system, ${\cal H}$ denoting the Hilbert space of the two
identical systems, for finite dimension $d=\operatorname{dim}({\cal
H})$. As well known \cite{Holevo}, the POVM of the measurement has the
form
\begin{eqnarray}
\d g \, P_g=\d g \, U_g \xi U_g^\dag\,,
\end{eqnarray}
where $\xi$ is a suitable positive operator such that the POVM is
normalized, and $\d g$ denotes a (suitably normalized) Haar invariant
measure on the group (for what we need in the following, the group is
unimodular, which guarantees that an invariant measure always
exists). For later convenience, we will normalize the Haar measure
over the group as $\int_{\bf G}\d g=d$ ($d$ is the dimension of the
Hilbert space on which the group is represented).  For infocomplete
measurements every operator can be expanded over the POVM density,
namely $P_g$ spans the whole linear space of linear operators.
Clearly, all expansions for bounded operators must be square summable
over the group ${\bf G}$. However, since we are considering only
finite dimensions, all operators are bounded, and since we consider
only compact groups (which then admit normalizable invariant Haar
measure), all group integrals are bounded too. In the present case,
the noise in Eq. (\ref{deltorho}) can be rewritten as
\begin{equation}
\delta O^2 (\rho ;P)
\doteq \int _{\bf G} \d g |\Tr[D^\dag_g O]|^2\Tr[P_g\rho]- |\Tr[O\rho]|^2\,.
\label{deltorho2}
\end{equation}
We will consider the following classes of pure states with uniform
prior probability distribution: $a$) all pure input states; $f$) all
factorized pure input states; $e$) all maximally entangled input
states (also called EPR states).  The averaged noise in all three
cases can be evaluated with the following integrals
\begin{eqnarray}
\delta_a O^2[P]&=&\frac1{d^2}\int_{\SU {d^2}}\!\!\!\!\!\d h\;\delta
O^2 \left (U_h|0\kk\bb 0|U_h^\dag;P \right )\;,
\label{deltall}\\ \delta_f
O^2[P]&=&\frac1{d^2}\int_{\SU d}\!\!\!\!\!\d h\int_{\SU d}
\!\!\!\!\!\d h^\prime\;\delta O^2 \left((V_h\otimes V_{h^\prime})
|0\>\<0|^{\otimes2}(V_h^\dag\otimes V_{h^\prime}^\dag);P \right )
\;,\label{deltfact}\\
\delta_e O^2[P]&=&\frac1d\int_{\SU {d}}\!\!\!\!\!\d h\;\delta O^2\left
  ((V_h\otimes I){\cal I}(V_h^\dag\otimes I);P\right )\label{deltbell}\;,
\end{eqnarray}
where ${\cal I}=\frac 1d \dket{I}\dbra{I}$, while $|0\kk$ and $|0\>$
are arbitrary reference pure states in $\sH^{\otimes2}$ and $\sH$,
respectively. Notice that the average of $|\<O\>|^2\equiv
|\Tr[O\rho]|^2$ over input states does not depend on the POVM, but
only on the set of states over which the average is estimated. Upon
denoting by overline the average over input states, one has
\begin{eqnarray}
\overline{|\<O\>|^2_a}& \doteq&\frac1{d^2}\int_{\SU {d^2}}\!\!\!\!\!\d
h\; \Tr \left [U_h^{\otimes 2}|0\kk\bb
0|^{\otimes2} U_h^{\dag \otimes 2}(O\otimes O^\dag )\right ] \;,\\
\overline{|\<O\>|^2_f}&\doteq&\frac1{d^2}\int_{\SU d}\!\!\!\!\!  \d
h\int_{\SU d}\!\!\!\!\!\d h^\prime\;\Tr \left [(V_h\otimes
V_{h^\prime})^{\otimes2}|0\>\<0|^{\otimes4}({V_h^\dag}\otimes
{V_{h^\prime}^\dag})^{\otimes2}(O\otimes O^\dag )\right ]\;,\\
\overline{|\<O\>|^2_e}&\doteq& \frac1d\int_{\SU {d}}\!\!\!\!\!\d
h\;\Tr\left [(V_h\otimes I)^{\otimes2}{\cal I}^{\otimes2}(V_h^\dag\otimes
I)^{\otimes2}(O\otimes O^\dag )\right ]\;,
\end{eqnarray}
These integrals can be evaluated by exploiting the following
identities (which are corollaries of Schur's lemmas)
\begin{eqnarray}
&&\int_{\SU d}\!\!\!\!\!\d g \,U_g X
U_g^\dag=\Tr[X]I\label{eq:gravtrc}\;,\\ &&\int_{\SU d}\!\!\!\!\!\d g
\,U_g^{\otimes 2}X U_g^{\dag \otimes 2}=\frac2{d+1}\Tr \left [P^{\sH}_S
X\right ]P^{\sH}_S+\frac2{d-1}\Tr\left [P^{\sH}_A X\right ]P^{\sH}_A\label{eq:grav2}\;,
\end{eqnarray}
where $P^{\sK}_S$ and $P^{\sK}_A$ denote the projection on the
symmetric and antisymmetric subspaces of ${\sK}^{\otimes2}$,
respectively. The result is
\begin{eqnarray}
&&
\overline{|\<O\>|^2_a}=\frac2{d^2(d^2+1)}\Tr\left [P^{\sH^{\otimes2}}_S(O\otimes
O^\dag )\right ] \;, \nonumber\\ &&
\overline{|\<O\>|^2_f}=\frac4{d^2(d+1)^2}\Tr\left [({P^{\sH}_S}_{13}\otimes
{P_S^{\sH}}_{24})(O\otimes O^\dag )\right ] \;,\nonumber\\ &&
\overline{|\<O\>|^2_e}=\frac2{d^3(d+1)}\Tr\left [({P_S^{\sH}}_{13}\otimes
{P_S^{\sH}}_{24})(O\otimes O^\dag )\right ]\;,
\end{eqnarray}
where $X_{ij}$ denotes an operator acting on ${\sH}_i \otimes
{\sH}_j$.  Finally, recalling that $P_S=\frac12(I+E)$ where $E$ is the
swap operator $E|\phi\>|\psi\>=|\psi\>|\phi\>$, and that
$\Tr[E(A\otimes B)]=\Tr[AB]$, we get
\begin{eqnarray}
&&\overline{|\<O\>|^2_a}=\frac1{d^2(d^2+1)}\left
  (\Tr \left [|O|^2\right]+|\Tr[O]|^2 \right )\label{avoall} \;,\\
  &&\overline{|\<O\>|^2_f}=
  \frac1{d^2(d+1)^2}\left(\Tr\left [ |O|^2 \right]+
|\Tr[O]|^2+ \Tr\left[|\Tr_1[O]|^2+ |\Tr_2[O]|^2 \right ]\right )
\label{avofact}\;,\\
  &&\overline{|\<O\>|^2_e}=\frac1{2d^3(d+1)}\left(\Tr\left [|O|^2\right]+
|\Tr[O]|^2
  + \Tr\left[|\Tr_1[O]|^2+|\Tr_2[O]|^2\right ]\right )\label{avobell}\;,
\end{eqnarray}
Notice that
$\overline{|\<O\>|^2_e}=\frac{d+1}{2d}\overline{|\<O\>|^2_f}$.

\par One can easily show that the first integral in Eq. (\ref{deltorho2})
is independent of the input state ensemble, so that the first term of
the noise depends only on the POVM, whereas the second term depends
only on the input ensemble.  This can be verified using identities
(\ref{eq:gravtrc}) and (\ref{eq:grav2}). One has
\begin{equation}
\delta O^2_x[P]=\frac1{d^2}\int_{\bf G}\d g|\Tr[D^\dag_g
O]|^2\Tr[P_g]-\overline{|\<O\>|^2_x}\,,
\label{genrum}
\end{equation}
where $x=a,f,e$, and a dual that optimizes one of these noise
parameters optimizes all of them. According to Sect.
\ref{s:processing}, we are in the situation where the canonical dual
is optimal, namely it provides the optimal processing function
minimizing the noise (\ref{deltorho2}). In the following section we
will then consider processing functions obtained from the canonical
dual.

\section{Product of local infocomplete measurements}\label{s:factorized}

As the first example of infocomplete POVM we consider
\begin{eqnarray}
P^{\mathrm{loc}}_{g,h}= U_g \nu U_g^\dag 
\otimes U_h \nu' U_h^\dag\label{lok}
\;,
\end{eqnarray}
where the elements $g,h$ belong to $\SU d$, and $\nu \,, \nu '$ are
pure states in ${\mathbb C}^d$. Such a POVM describes a measurement that
can be performed locally by two separate parties, with classical
communication needed in order to evaluate the processing function.

It can be easily shown by Schur's lemma that $P_g\otimes P_h$ is
actually a POVM. The canonical dual can be written as \cite{ic1}
\begin{eqnarray} 
D_{g,h}= [(d+1)U_g \nu U_g^\dag-I ]\otimes [(d+1)U_h \nu ' U_h^\dag-I
 ] \;.
\end{eqnarray}
The noise $\delta_x O^2[P^{\mathrm{loc}}]$ in the evaluation of the
expectation value of an operator $O$ is given by
\begin{equation}
\delta_x O^2[P^{\mathrm{loc}}]=\frac1{d^2}\int_{\SU d}\!\!\!\!\!  \d
g\int_{\SU d}\!\!\!\!\!\d
h|\Tr[D^\dag_{g,h}O]|^2-\overline{|\<O\>|_x^2}\,.
\label{eq:avpur}
\end{equation}
Substituting the expression for the canonical dual and exploiting the
identities in Eqs.~(\ref{eq:gravtrc}) and (\ref{eq:grav2}), we obtain
\begin{eqnarray}
\delta_x O^2 [P^{\mathrm{loc}}] &=& 
\frac1{d^2}\left\{(d+1)^2\Tr \left [|O|^2 \right ] +|\Tr[O]|^2 \right. 
\nonumber \\&&-
\left.  (d+1)\Tr \left [|\Tr _1
[O]|^2+|\Tr _2 [O]|^2\right ]\right\}-\overline{|\<O\>|^2_x}\;.
\end{eqnarray}

\section{Global infocomplete measurement}\label{s:global}

We will now consider the POVM
\begin{eqnarray}
P^{\mathrm{glob}} _g = U_g \nu U^\dag_g\;,
\end{eqnarray}
where $g$ now belongs to $\SU{d^2}$, $\nu $ is a pure state in
${\mathbb C}^d \otimes {\mathbb C}^d$, and $\int_{\SU{d^2}}\d g=d^2$.
The canonical dual set is given by \cite{ic1}
\begin{eqnarray}
D_g= (d^2 +1) U_g \nu U_g ^\dag- I\;.
\end{eqnarray}
The average noise over all pure states can be evaluated as in the
previous section, and one obtains 
\begin{eqnarray}
\delta O^2_x[P^\mathrm{glob}]= \frac1{d^2}\left\{(d^2+1)\Tr [|O|^2]
  -|\Tr[O]|^2\right\}-
\overline{|\<O\>|^2_x}\;.
\end{eqnarray}

\section{Bell measurement}\label{s:Bell}
Finally, we will consider the Bell POVM
\begin{eqnarray}
P^\mathrm{Bell}_g = U_g \otimes I \,\dket{I}\dbra{I} \, U^\dag_g \otimes I
= d (U_g \otimes I )\,{\cal I}\, (U^\dag_g \otimes I)
\;.\label{eq:bell}
\end{eqnarray}
with $g$ belonging to $\SU d$. Using Eq. (\ref{eq:gravtrc}) it is easy
to verify that Eq. (\ref{eq:bell}) actually defines a POVM. On the
other hand, such a POVM is not infocomplete since operators whose
partial trace is not proportional to the identity cannot be spanned.
This can be seen directly from the reconstruction formula
(\ref{eq:reconst}).  We can evaluate explicitly the frame operator as
follows
\begin{eqnarray}
F&=&\int_{\SU d}\d g \dket{P_g}\dbra{P_g}\nonumber\\
&=& d^2\int_{\SU d}\d g\ U_g\otimes I\otimes U_g^{*}\otimes I(
{\cal I}_{12}\otimes{\cal I}_{34}) U_g^{\dag}\otimes I\otimes U_g^{\tau}\otimes I
\nonumber \\
& = &d \left \{ {\cal I}_{13}\otimes {\cal I}_{24} +
\frac{1}{d^2 -1}(I_{13}-{\cal I}_{13})
\otimes (I_{24}-{\cal I}_{24})
\right \}\;,
\end{eqnarray}
where ${\cal I}_{ij}$ denotes the projector $\cal I$ acting on
$\sH_i\otimes\sH_j$. Indeed, $F$ has a nontrivial kernel. The Bell
POVM in Eq. (\ref{eq:bell}) can be used, however, to obtain the
expectation values of any operator of the form
\begin{eqnarray}
O= \alpha _{00} I +\sum_{i,j\neq 0} \alpha _{ij} V_i \otimes W_j\;,\label{ouv}
\end{eqnarray}
where $V_i$ and $W_j$ are orthogonal bases for the operators with
$V_0=W_0=I$ (and hence with $V_i$ and $W_j$ traceless for $i,j \neq
0$).  These are exactly the operators in the support of $F$. Notice
also that any maximally entangled state is of the form (\ref{ouv}).
\par Since one has
\begin{eqnarray}
F^\ddag\dket{U_g}_{12} \dket{U_g^*}_{34}
=\frac{d^2 -1}{d}\dket{U_g}_{12} \dket{U_g^*}_{34} -\frac{d^2 -2}{d^2}
\dket{I}_{13} \dket{I}_{24}
\;,
\end{eqnarray}
the canonical dual is given by
\begin{eqnarray}
D_g= \frac{d^2 -1}{d}\dket{U_g} \dbra{U_g} -\frac{d^2 -2}{d^2 } I
\;.
\end{eqnarray}
As long as operators $O$ of the form (\ref{ouv}) are considered, the
average noise can still be evaluated through Eq.~(\ref{genrum}), and
is given by
\begin{eqnarray}
\delta O^2_x[P^\mathrm{Bell}]=\frac1{d}\int_{\SU d}\d g \, |\Tr[D_g
O]|^2-\overline{|\<O\>|_x^2}\,.
\end{eqnarray}
Evaluation of the integral gives the following result
\begin{eqnarray}
\delta O^2_x[P^\mathrm{Bell}] &=&
\frac1{d^2}\left\{(d^2-1)\Tr[|O|^2]+|\Tr[O]|^2\right\} \nonumber \\&&
-\left(\frac{d^2-1}{d^3}\right) 
\Tr \left [|\Tr_1[O]|^2+|\Tr_2[O]|^2 \right ]-\overline{|\<O\>|^2_x}\,.
\end{eqnarray}

\section{Comparison and conclusions}\label{s:comparison}
We now compare the three infocomplete POVM's in terms of their
statistical efficiency, namely we compare their variance
(\ref{deltorho2}). We have
\begin{eqnarray}
\delta O^2_x[P^\mathrm{glob}]-\delta O^2_x[P^\mathrm{Bell}]&=&
\frac{2}{d^2}(\Tr[|O|^2]-|\Tr[O]|^2) 
+\frac{d^2-1}{d^3}\Tr \left 
[|\Tr_1[O]|^2+|\Tr_2[O]|^2 \right ]\,, \nonumber \\
\delta O^2_x[P^\mathrm{loc}]-
\delta O^2_x[P^\mathrm{glob}]
&=&
\frac{2}{d}(\Tr[|O|^2]+\frac{1}{d}|\Tr[O]|^2) 
-\frac{d+1}{d^2} \Tr \left [|\Tr_1[O]|^2+|\Tr_2[O]|^2\right ]\,, \nonumber \\
\delta O^2_x[P^\mathrm{loc}]-\delta O^2_x[P^\mathrm{Bell}]&=&
\frac{2(d+1)}{d^2}\Tr[|O|^2]
-\frac{d+1}{d^3}\Tr\left [|\Tr_1[O]|^2+|\Tr_2[O]|^2\right ]\,.
\end{eqnarray}
Clearly, for operators $O$ having $\Tr_1[O]=\Tr_2[O]=0$ (e.~g. for
operators of the form $O=A\otimes B$ with both $A$ and $B$ traceless)
one has
\begin{eqnarray}\label{gencomparison}
\delta ^2 O_x[P^\mathrm{Bell}] \leq\delta ^2
O_x[P^\mathrm{glob}]\leq\delta ^2 O_x[P^\mathrm{loc}] \;.
\end{eqnarray}
For off-diagonal matrix elements in a factorized basis, namely with
$O=|i\>\<j|\otimes|n\>\<m|$ and $i\neq j$ and/or $n\neq m$, the
inequalities (\ref{gencomparison}) also hold true, and, specifically,
for diagonal matrix elements one has $\delta ^2
O_x[P^\mathrm{glob}]=\delta ^2 O_x[P^\mathrm{loc}]$. Also for
maximally entangled projectors $O=\frac 1d \dket{V}\dbra{V}$ the
variances are bounded as in Eq. (\ref{gencomparison}), whereas for
$O=\frac 1d \dket{V}\dbra{W}$ and $\Tr[W^\dag V]=0$ one has
\begin{eqnarray}
\delta ^2 O_x[P^\mathrm{glob}]\leq\delta ^2 O_x[P^\mathrm{loc}]\;,
\end{eqnarray}
and generally the expectation of $O$ cannot be estimated using the
Bell POVM, unless $W^\dag V\propto I$. This means that for matrix
elements in an orthogonal Bell basis $\dket{V_n}$ where $V_n$ are
traceless (apart from the identity) and form a also a (projective)
representation of a group (such as the usual orthonormal basis with
Pauli matrices for $d=2$, or shift-and-multiply operators in dimension
$d>2$), the Bell POVM is optimal for diagonal matrix elements, whereas
for off-diagonal matrix elements it cannot be used, and for these it
is better to use the global measurement than the local one.

We conclude that for factorized traceless operators, for Bell
projectors, and for the density matrix estimation in factorized basis
the Bell measurement always performs better than the unrestricted
global measurement, which in turn outperforms the local one. For the
diagonal matrix elements in the factorized basis the Bell measurement
cannot be used, and global and local POVM's perform equally
well. Finally, for matrix elements in an orthogonal Bell basis the
Bell POVM is optimal for diagonal matrix elements, whereas for
off-diagonal matrix elements it cannot be used, and for these elements
the global measurement performs better than the local one.

A further interesting study continuing the present analysis is the
evaluation of the noise for infocomplete measurements with minimal
number of outcomes, comparing local versus global measurements.

\section*{Acknowledgments}
This work has been sponsored by INFM through the project
PRA-2002-CLON, and by EC and MIUR through the cosponsored ATESIT
project IST-2000-29681 and Cofinanziamento 2003. The authors thank
Susana Huelga for stimulating discussions and Alfredo G\'omez-Rodr\'\i
guez for his useful suggestions.


\begin{thebibliography}{99}
\bibitem{corr}Equation (\ref{deltorho}) corresponds to the trace of
  the covariance matrix of statistical errors. It becomes the usual
  expression of the variance of the estimation of an observable for
  Hermitian operators $O$.
\end{thebibliography}
\end{document}